\documentclass[prl,twocolumn,showpacs,floatfix]{revtex4-1}
\usepackage{amsmath}
\usepackage{amssymb}
\usepackage{graphicx}
\usepackage[english]{babel}
\usepackage{latexsym}
\usepackage{graphics}
\usepackage{subfigure}
\begin{document}
\title{Quantization scheme of surface plasma polariton in helical liquid 
and the exchanging interaction between quasi particles and emitters}
\author{Li Mao}
\thanks{Email: maoli@whu.edu.cn}
\author{Hongxing Xu}
\thanks{Email: hxxu@whu.edu.cn}
\begin{abstract}
The collective modes of helical electron gases 
interacting with light have been studied in an extended random phase 
approximation. By separating two kinds of electron density oscillations, 
the complicate operator dynamics coupling electrons and photons can be 
simplified and solved. The inverse operator transformation that 
interprets electron oscillations and photons with quasi particles has 
been developed to study the interaction between surface plasma 
polaritons (SPPs) and emitters. Besides the ordinary interaction induced 
by electric field, we find an additional term which plays important 
roles at small distance arising from electron exchanging effect.
\end{abstract}
\affiliation{School of Physics and Technology, Wuhan University,
Wuhan, Hubei, 430072 CN} 
\maketitle
\emph{Introduction}--Recently, the quantum nature of surface plasmon 
poloritons (SPPs) \cite{tame_quantum_2013}, the coupled electromagnetic 
waves (or photons) and electron collective oscillations (or plasmons), 
has attracted great interests. It is well known that the electric fields 
of SPPs are quite localized and strongly enhanced \cite{raether_surface_1988}, 
so, it provides an excellent platform for investigating light matter interaction 
\cite{chang_quantum_2006}. With the remarkable progresses in theories 
and device fabrications, people now can study and analyze the light 
matter interaction in the extreme spatial confinement
\cite{xu_spectroscopy_1999,chen_probing_2018}, where the quantum effect becomes 
significant, such as the quantum tunneling 
\cite{danckwerts_optical_2007,zuloaga_quantum_2009,mao_effects_2009} and 
size effects \cite{scholl_quantum_2012}. Also, it has been shown that 
SPPs can conserve the energy-time entanglement of a pair of photons 
\cite{fasel_energy-time_2005}, and can be used in integrated logic devices 
\cite{wei_quantum_2011,wei_cascaded_2011}. And 
due to the strong coupling between SPPs and quantum emitters, such 
hybrid systems can be used as single photon sources 
\cite{de_leon_tailoring_2012}. 

Due to the novel properties of Dirac electrons\cite{qi_topological_2011}, 
graphene and topological insulator (TI) become very hot topics in condensed 
matter physics. Most recently, plasmons and SPPs are proposed and 
confirmed to exist in graphene\cite{mikhailov_new_2007,hwang_dielectric_2007,grigorenko_graphene_2012} 
which provides highly tunable plasmonic metamaterials\cite{ju_graphene_2011}, 
and in TI surface\cite{pietro_observation_2013} which can be considered as coupled 
electron density oscillation and spin density 
oscillation\cite{raghu_collective_2010}. People also find direct evidence 
for three dimensional (3D) Dirac plasmon in the type-II Dirac 
semimetal\cite{politano_3d_2018}.  

Generally, SPPs are regarded as quasi particles which hybrid photons and
electron, collective oscillations. The quantization scheme of SPPs plays 
a fundamental role when dealing with the interaction between them and 
other particles or quasi-particles\cite{gonzalez-tudela_theory_2013}, 
also, it is important to the plasmon 
mediated interactions\cite{bill_electronic_2003}. When the momentum is not quite small, 
one can ignore photons and the quantization can be achieved by 
directly writing down the plasmon operator as the summation of electron 
density oscillating operators, then the wave function and energy are 
solved with random phase approximation (RPA) \cite{brout_correlation_1957, 
efimkin_collective_2012,raghu_collective_2010}. In such method, the 
interaction between electrons is the static Coulomb interaction, so the 
retardation effect\cite{kukushkin_observation_2003} is totally ignored. 
While the another scheme developed 
from quantizing the electromagnetic wave \cite{tame_quantum_2013, 
gruner_green-function_1996}, which is widely applied, needs to solve 
Maxwell's equations with the materials' dielectric functions. Which means, 
in this method, the roles of electrons in SPPs are hidden behind the 
dielectric functions.

\begin{figure}[]
\includegraphics[width=0.9\linewidth]{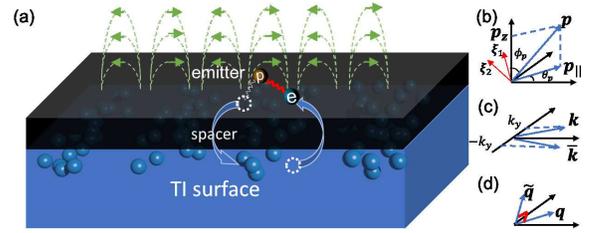} \vspace{-0pt}
\caption{(Color online) (a). A molecular emitter (e means electron and p 
means positively charged core) is placed on top of a topological insulator 
with a spacer layer. When SPPs are excited in the TI surface, they can 
coupled to each other by the electromagnetic field and through exchanging 
electrons. (b). The coordinate system used in this paper. (c). The 
illustration of $\bar{\mathbf{k}}=(k_x,-k_y,0)$. (d). The illustration of 
$\tilde{\mathbf{q}}$ which perpendicular to $\mathbf{q}$ with 
$\theta_{\tilde{\mathbf{q}}}=\theta{\mathbf{q}}+\pi/2$.}
\end{figure}

To clarify all components' roles in SPPs, we extend the first method 
\cite{efimkin_collective_2012,raghu_collective_2010} to include both photons 
and electrons. Firstly, we start from the full electron light interaction 
Hamiltonian, describe SPPs as the compositions of electron oscillations and 
photons. After separating two kinds of electron density oscillation 
operators, we get the quasi particle energies and wave functions 
with an extended RPA method. The wave functions are proved to satisfy a 
Schr\"odinger like equation of a none Hermitian Hamiltonian, and 
have modified orthogonal conditions. Once getting these orthogonal conditions, 
we can calculate the inverse operator transformation which expresses 
electron oscillations and photons with quasi particles. Finally, with 
the help of inverse operator transformation, we can calculate the Coulomb 
field and vector field and analyze each component's contribution to the 
total electric field. Also, one can calculate the interaction 
between SPPs and other particles, as an example, we've calculated the 
SPP/emitter interaction and find it contains two parts, the first one is 
just $\hat{\mathbf{\mu}}\cdot\hat{\mathbf{E}}$ term offered by the second scheme 
\cite{archambault_quantum_2010,gonzalez-tudela_theory_2013}, the second one 
is an additional interaction induced by exchanging electrons between them  
(The physical picture is illustrated in Figure 1a, where emitter electron 
can jump to 2DEG and join the oscillation of SPPs then go back to emitter 
at some other place), which has different selection rules and comparable 
strength at large q, and plays the most important role at small momentum.

\emph{Quantization Scheme of SPPs}--The system we have considered can be 
described by Figure 1a, where a emitter (2D exciton or molecule) is 
closely placed on top of a 2D electron gas. Note that we've chosen the 
surface state of topological insulator (TI) \cite{qi_topological_2011}, 
and it can be straightforwardly extended to others systems, such as 
graphene, single layer MoS$_2$, or quantum wells formed by semiconductor 
films. In the 2DEGs, SPPs can generate and propagate and interact with 
the emitter through electromagnetic field or through electron exchanging 
effect. 

Firstly, let's ignore the emitter and calculate SPPs of the 2DEGs. 
Generally, the energies of SPPs we've considered in this paper are quite 
small compared to the band gap of 2DEG. So, an effective low energy 
Hamiltonian, $\hat{H}_s(\hat{\mathbf{p}})=\hat{\mathbf{p}}^2/2m^*
+\hbar v_f(\hat{\sigma}_x\hat{p}_y-\hat{\sigma}_y\hat{p}_x)$ for TI 
surface states, is sufficient to describing electrons. Note that, the 
topological trivial term $\hat{\mathbf{p}}^2/2m^*$ is essential to 
provide a diamagnetic current, which overcomes the energy cutoff problem 
when dealing with the optical properties of TI surface states 
\cite{stauber_dynamical_2010}, and also, we find this term renormalizes 
the paramagnetic current-current response function to the physical one 
(For more details, please see the supplementary material.). Under these 
considerations, the total electron light interaction Hamiltonian in 
Heisenberg picture reads (Coulomb gauge $\nabla \cdot \hat{\mathbf{A}}=0$ 
has been used in this paper).
\begin{align}
&\hat{\mathbf{H}}(t)=\int d\mathbf{x}\hat{\Psi}^{\dag}(\mathbf{x}
  ,t)H_s[\hat{\mathbf{p}}+\frac{e}{c}\hat{\mathbf{A}}(\mathbf{x},t)]
   \hat{\Psi}(\mathbf{x},t)\nonumber\\
+&\frac{1}{2}\int d\mathbf{x}d\mathbf{x}'\hat{\Psi}^{\dag}
   (\mathbf{x},t)\hat{\Psi}^{\dag}(\mathbf{x}',t')
   V(|\mathbf{x}-\mathbf{x}'|)\hat{\Psi}(\mathbf{x}',t)
   \hat{\Psi}(\mathbf{x},t)\nonumber\\
+&\sum_{\lambda\mathbf{p}}\hbar \omega_{\mathbf{p}}
   \hat{a}^{\dag}_{\lambda\mathbf{p}}(t)\hat{a}_{\lambda\mathbf{p}}(t),
\end{align}
where $\hbar\omega_{\mathbf{p}}$ is the free photon energy, and 
$\mathbf{x}$ and $\mathbf{p}$ are the 3D position and momentum 
vector respectively. In this paper, we have used cylinder coordinates 
that $\mathbf{x}=(\mathbf{r},z)$ and $\mathbf{p}=(\mathbf{p}_{\parallel}
,p_z)$. The photon index $\lambda=1,2$ corresponding to two photon modes 
with the polarization vectors $\mathbf{\xi}_{\lambda\mathbf{p}}$ 
perpendicular to the wave vector $\mathbf{p}$, which have been set to 
$\mathbf{\xi}_{1}(\mathbf{p})=(-\sin{\theta_{\mathbf{p}}},
\cos{\theta_{\mathbf{p}}},0)$ and $\mathbf{\xi}_{2}(\mathbf{p})=
(-\cos{\phi_{\mathbf{p}}}\cos{\theta_{\mathbf{p}}},-\cos{\phi_{\mathbf{p}}}
\sin{\theta_{\mathbf{p}}},\sin{\phi_{\mathbf{p}}})$, where $\theta_{\mathbf{p}}$ 
and $\phi_{\mathbf{p}}$ are angles of $\mathbf{p}$ in polar coordinate 
(please see Figure 1b). The radiation field $1/c\hat{\mathbf{A}}
(\mathbf{x},t)=1/\sqrt{\nu}\sum_{\lambda\mathbf{p}}e^{i\mathbf{p}\cdot\mathbf{x}}
\hat{\mathbf{A}}_{\lambda\mathbf{p}}(t)$, here $\nu=SL_z$ is the space volume, and 
$\hat{\mathbf{A}}_{\lambda\mathbf{p}}(t)=\sqrt{2\pi\hbar/
  \omega_{\mathbf{p}}}[\mathbf{\xi}_{\lambda}(\mathbf{p})
  \hat{a}_{\lambda\mathbf{p}}(t)+\mathbf{\xi}_{\lambda}
  (-\mathbf{p})\hat{a}^{\dag}_{\lambda,-\mathbf{p}}(t)]$, 
where $\hat{a}_{\lambda\mathbf{p}}(t),\ \hat{a}^{\dag}_{\lambda,
-\mathbf{p}}(t)$ are photon annihilate and creation operators. 

Aware that the electron excitation energy in z direction is quite bigger 
then the energy of plane excitations we've considered, so, one can 
simplify the real space operator $\hat{\Psi}^{\dag}(\mathbf{x},t)$ 
with $\psi(z)\hat{\Psi}^{\dag}(\mathbf{r},t)$. Where $\psi(z)$ is the 
envelope function ($\int|\psi(z)|^2dz=1$), and $\hat{\Psi}(\mathbf{r},t)
=(\hat{c}_{\uparrow}(\mathbf{r},t), \hat{c}_{\downarrow}(\mathbf{r},t))^T$ 
is a Nambu spinor. In momentum space, the single electron Hamiltonian 
$\hat{H}_s$ can be diagonalized by transformation 
$\hat{\gamma}_{s\mathbf{k}}=\langle s\mathbf{k}|\hat{\Psi}_{\mathbf{k}}$, 
where $s=\pm 1$, $\langle s\mathbf{k}|=e^{-\frac{i}{2}s
\theta_{\mathbf{k}}}(u_{sk}e^{\frac{i}{2}\theta_{\mathbf{k}}},-isv_{sk}
e^{-\frac{i}{2}\theta_{\mathbf{k}}})$ and $\hat{\Psi}_{\mathbf{k}}=
(\hat{c}_{\mathbf{k}\uparrow},\hat{c}_{\mathbf{k}\downarrow})^T$ with 
energy $\xi_{sk}=\mu+sE_k$, $E_k=\sqrt{v_f^2k^2+h^2}$, and wave 
functions $u_{sk},v_{sk}=\sqrt{\frac{1}{2}(1\pm sh/E_k)}$. For 
shorter notations, let's omit the subscript of $\mathbf{k}_{\parallel},
\mathbf{q}_{\parallel}$ and let $\mathbf{k},\mathbf{q}$ just be the 
in-plane vectors, omit time $t$ in operators for the same 
reason, and define $W^{ss'}_{\mathbf{k}\mathbf{q}}=\langle 
s\mathbf{k}+\mathbf{q}|s'\mathbf{k}\rangle$, 
$X^{ss'}_{\mathbf{k}\mathbf{q}}=\langle s\mathbf{k}+\mathbf{q}|
\sigma_x|s'\mathbf{k}\rangle$, $Y^{ss'}_{\mathbf{k}\mathbf{q}}
=\langle s\mathbf{k}+\mathbf{q}|\sigma_y|s'\mathbf{k}\rangle$, 
and $\Delta \xi^{ss'}_{\mathbf{k}\mathbf{q}}=\xi_{s\mathbf{k}
+\mathbf{q}}-\xi_{s'\mathbf{k}}$, $\Delta n^{ss'}_{\mathbf{k}\mathbf{q}}
=n_{s'\mathbf{k}}-n_{s\mathbf{k}+\mathbf{q}}$. Finally, the Hamiltonian 
in momentum space can be expressed as 
\begin{align}
\label{eq:Hk}
&\hat{\mathbf{H}}=\sum_{s\mathbf{k}}\xi_{s\mathbf{k}}
\hat{\gamma}_{s\mathbf{k}}^{\dag}\hat{\gamma}_{s\mathbf{k}}+
\frac{1}{2S}\sum_{ss'll'\mathbf{k}\mathbf{k}'\mathbf{q}} 
W^{ss'}_{\mathbf{k}\mathbf{q}}W^{ll'}_{\mathbf{k}',-\mathbf{q}}
V_{q}\hat{\gamma}_{s\mathbf{k}+\mathbf{q}}^{\dag}\nonumber\\
&\hat{\gamma}_{l\mathbf{k}'-\mathbf{q}}^{\dag}
\hat{\gamma}_{l'\mathbf{k}'}\hat{\gamma}_{s'\mathbf{k}}
+\sum_{\lambda\mathbf{q}}\hbar \omega_{\mathbf{q}}
\hat{a}^{\dag}_{\lambda\mathbf{q}}\hat{a}_{\lambda\mathbf{q}}
+\frac{ev_f}{\sqrt{\nu}}\sum_{\mathbf{k}\mathbf{q}ss'}\hat{\gamma}_
{s\mathbf{k}+\mathbf{q}}^{\dag}\hat{\gamma}_{s'\mathbf{k}}\nonumber\\
&
[Y^{ss'}_{\mathbf{k}\mathbf{q}}\hat{A}^{x}_{\mathbf{q}}
-X^{ss'}_{\mathbf{k}\mathbf{q}}\hat{A}^{y}_{\mathbf{q}}]
+\frac{\Pi^d}{2L_z}\sum_{\mathbf{q}p_z',l=x,y}\hat{A}^{l}_{\mathbf{q}}
\hat{A}^l_{(-\mathbf{q},-p_z')},
\end{align}
where $V_{q}=2\pi e^2/q$ is the two dimensional Coulomb potential, the 
last term provides the diamagnetic current arising from the topological 
trivial term in $\hat{H}_s$ mentioned previously. 

In order to get the collective excitations or SPPs, one can always 
define an operator compositing of electron oscillations and photons 
($\hat{\rho}^{\dag}_{\mathbf{k}\mathbf{q}ss'}=\hat{\gamma}^
{\dag}_{\mathbf{k}+\mathbf{q}s}\hat{\gamma}_{\mathbf{k}s'}$, 
$\hat{a}_{\lambda\mathbf{p}}$, and $\hat{a}^{\dag}_{\lambda\mathbf{p}}$), 
and may find all of them coupled to each other and make the dynamics 
quite complicate. The key point to simplify our calculations is defining 
two kinds of charge oscillation operators $\hat{\rho}^{l\dag}_{\mathbf{k}
\mathbf{q}ss'}$ and $\hat{\rho}^{t\dag}_{\mathbf{k}\mathbf{q}ss'}$ that 
\begin{align}
\begin{pmatrix}
 \hat{\rho}^{l\dag}_{\mathbf{k}\mathbf{q}ss'}\\
 \hat{\rho}^{t\dag}_{\mathbf{k}\mathbf{q}ss'}
\end{pmatrix}
=\frac{1}{\sqrt{2}|W^{ss'}_{\mathbf{k}\mathbf{q}}|}
\begin{pmatrix}
 W^{ss'}_{\mathbf{k}\mathbf{q}} & W^{ss'}_{\bar{\mathbf{k}}\mathbf{q}}\\
W^{ss'}_{\mathbf{k}\mathbf{q}} & -W^{ss'}_{\bar{\mathbf{k}}\mathbf{q}}
\end{pmatrix}
\begin{pmatrix}
\hat{\rho}^{\dag}_{\mathbf{k}\mathbf{q}ss'}\\
\hat{\rho}^{\dag}_{\bar{\mathbf{k}}\mathbf{q}ss'},
\end{pmatrix}
\end{align}
where $\bar{\mathbf{k}}=(k_x,-k_y,0)$ (illustrated in Figure 1(c)). The 
upper script $l$ means longitudinal and $t$ means transverse, one can 
easily check the charge density $\hat{\rho}^{\dag}_{\mathbf{q}}$ and the 
longitudinal current density $\hat{j}^{x\dag}_{\mathbf{q}}$ can be 
written as the summation of $\hat{\rho}^{l\dag}_{\mathbf{k}\mathbf{q}ss'}$ 
while the transverse current density $\hat{j}^{y\dag}_{\mathbf{q}}$ 
can be composited by $\hat{\rho}^{t\dag}_{\mathbf{k}\mathbf{q}ss'}$. 
Note that $(W_{\mathbf{k}\mathbf{q}}^{ss'})^*= W^{ss'}_{\bar{\mathbf{k}}
\mathbf{q}}$, one immediately find $\hat{\rho}^{l\dag}_{ss'\mathbf{k}
\mathbf{q}}=\hat{\rho}^{l\dag}_{ss'\bar{\mathbf{k}}\mathbf{q}}$ and 
$\hat{\rho}^{t\dag}_{ss'\mathbf{k}\mathbf{q}}=-\hat{\rho}^{t\dag}
_{ss'\bar{\mathbf{k}}\mathbf{q}}$, which means we only need to consider 
half $\mathbf{k}$ space to make a complete set (we can set $k_{y}=\pm 
\pi/L_{y},\pm 3\pi/L_{y},...$ to avoid the special $k_y=0$ points). 

Similar to Nambu spinor, we can define a operator basis that $\hat{\Phi}
^{\dag}_{\mathbf{q}}=[\cdot\cdot\cdot\hat{\rho}^{\lambda_{\rho}\dag}_{
\mathbf{k}_i\mathbf{q}s_is_i'}\cdot\cdot\cdot\hat{a}_{\lambda_{a}
(-\mathbf{q},-p_{zi})}\cdot\cdot\cdot\hat{a}^{\dag}_{\lambda_{a}(\mathbf{q},
p_{zi})}\cdot\cdot\cdot]^T$, where $\lambda_{\rho}=l,t$, $\lambda_{a}=1,2$, 
$s_is_i'=++,+-,-+,--$ and the electron momentum $\mathbf{k}_i$ runs to 
every k points in the previously defined half space, the photon momentum 
$p_{zi}$ runs to all points in the z direction. Finally, the collective 
mode can be written as $\hat{Q}_{\mathbf{q}}^{\dag}=\Phi_{\mathbf{q}}
\hat{\Phi}^{\dag}_{\mathbf{q}}$, where $\Phi_{\mathbf{q}}$ is the wave 
function. By comparing the coefficients of different operators in 
the quasi particle dynamic equation $i\hbar\partial_t\hat{Q}_{\mathbf{q}}
^{\dag}=[\hat{Q}_{\mathbf{q}}^{\dag},\hat{H}]=-\hbar\Omega_{\mathbf{q}}
\hat{Q}_{\mathbf{q}}^{\dag}$, one can get the quasi particle energies 
($\hbar\Omega_{\mathbf{q}}$) and wave functions ($\Phi_{\mathbf{q}}$).
After some algebras, we find the quasi particle energies satisfy the 
following equation (for more details, please see the supplementary 
material.).
\begin{align}
V_{\mathbf{q}}^lV_{\mathbf{q}}^t|\Pi_{j_x j_y}|^2=(1-
V_{\mathbf{q}}^l\Pi_{j_xj_x})(1-V_{\mathbf{q}}^t\Pi_{j_yj_y})
\end{align}
where $V_{\mathbf{q}}^l$ and $V_{\mathbf{q}}^t$ are the longitudinal and 
transverse photon propagators, and $\Pi_{j_xj_x}$ $\Pi_{j_yj_y}$ are the 
diagonal current current response functions, $\Pi_{j_xj_y}$ is the off 
diagonal current current response function\cite{stauber_dynamical_2010,scholz_dynamical_2011}. 
We want to emphasize here that such equation is just the same one derived 
by traditional method. 

It is well known that the topological off diagonal current current 
response function is zero when the static Zeeman field $h=0$, which means 
the longitudinal and transverse modes are decoupled (Eq. (4) reduces 
to two independent equations $1-V^l_{\mathbf{q}}\Pi_{j_xj_x}(\Omega^l
_{\mathbf{q}},\mathbf{q})=0$, and $1-V^t_{\mathbf{q}}\Pi_{j_yj_y}
(\Omega^t_{\mathbf{q}},\mathbf{q})=0$). At the same time, the 
longitudinal and transverse modes can be expressed as $\hat{Q}_{\mathbf
{q}}^{l\dag}=\sum_{\mathbf{k}ss'}\Phi^{l}_{ss'\mathbf{k}\mathbf{q}}
\hat{\rho}^{l\dag}_{ss'\mathbf{k}\mathbf{q}}+\sum_{p_z}\Phi^{a}_{1p_z}
\hat{a}_{1(-\mathbf{q},-p_z)}+\Phi^{c}_{1p_z}\hat{a}^{\dag}_{1(\mathbf
{q},p_z)}$ and $\hat{Q}_{\mathbf{q}}^{t\dag}=\sum_{\mathbf{k}ss'}
\Phi^{t}_{ss'\mathbf{k}\mathbf{q}}\hat{\rho}^{t\dag}_{ss'\mathbf{k}
\mathbf{q}}+\sum_{p_z}\Phi^a_{2p_z}\hat{a}_{2(-\mathbf{q},-p_z)}+
\Phi^c_{2p_z}\hat{a}^{\dag}_{2(\mathbf{q},p_z)}$, where $\Phi^{l,t}_{ss'
\mathbf{k}\mathbf{q}}$ and $\Phi^{a,c}_{1,2p_z}$ are the corresponding 
components of wave function $\Phi_{\mathbf{q}}$ (As an example, we've 
plotted $|\Phi^{a,c}_{1p_z}|^2$ in Figure 2, and $|\Phi^{l}_{++\mathbf{k}
\mathbf{q}}|^2\Delta n^{++}_{\mathbf{k}\mathbf{q}}$ in the inset figure 
for $q=0.1k_f$). So, in this special case, the longitudinal density 
oscillation we've defined only coupling to the first photon mode, while 
the transverse density oscillation just coupling to the second photon 
mode. For general cases ($h\neq 0$), the longitudinal and transverse 
modes coupled to each other, and the details can be found in 
supplementary materials. 
\begin{figure}[]
\centering
\includegraphics[width=0.95\linewidth]{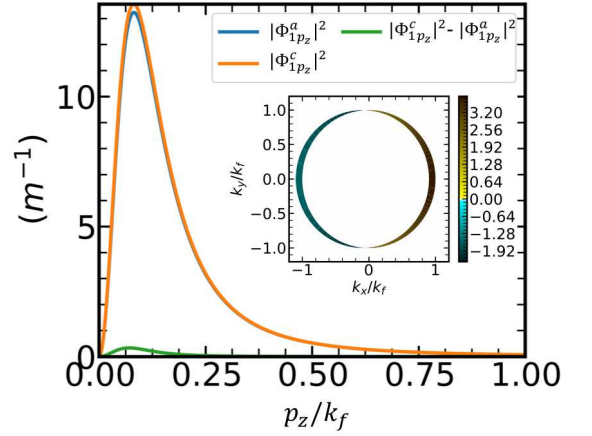} 
\caption{(Color online) $|\Phi^{a,c}_{1p_z}|^2$ and 
$|\Phi^{l}_{++\mathbf{k}\mathbf{q}}|^2\Delta n^{++}_{\mathbf{k}\mathbf{q}}$ 
(inset figure) as a function of $q_z$ and $\mathbf{k}$ at $q=0.1k_f$. 
The chemical potential for electron gas is $0.1eV$ and the Fermi velocity 
of TI $v_F=5\times 10^5m/s$.}
\end{figure}

The wave functions and energies can be proved to satisfy an eigenvalue 
equation that $\bar{H}\Phi_{n\mathbf{q}}=\hbar\Omega_{n\mathbf{q}}\Phi_
{n\mathbf{q}}$. In principle, one can always calculate and find all the 
solutions, then get $(\hat{Q}_{\mathbf{q}}^{\dag},\hat{Q}_{-\mathbf{q}},
\hat{Q}_{1\mathbf{q}}^{\dag},\hat{Q}_{1,-\mathbf{q}},\hat{Q}_{2\mathbf
{q}}^{\dag},\hat{Q}_{2,-\mathbf{q}},\cdot,\cdot,\cdot)^T=\bar{U}\hat
{\Phi}_{\mathbf{q}}^{\dag}$, the annihilator operators appear because 
$\bar{H}_{\mathbf{q}}$ has particle-hole symmetry that $R^{\dag}
\bar{H}_{\mathbf{q}}R=-\bar{H}^*_{-\mathbf{q}}$, $\hat{Q}_{-n,-\mathbf{q}}
=\hat{Q}^{\dag}_{n,\mathbf{q}}$, where $-n$ indicates negative energy. 
Note that $\bar{H}$ isn't Hermitian, instead, it satisfies 
$\bar{H}^{\dag}J=J\bar{H}$ with $J=diag[\Delta n^{s_1s'_1}_{\mathbf{k}
_1\mathbf{q}_{1}},\Delta n^{s_2s'_2}_{\mathbf{k}_2\mathbf{q}_2}\cdot\cdot
\cdot,1,1,\cdot\cdot\cdot,-1,-1,\cdot\cdot\cdot]$, one can immediately 
find $\langle\Phi_{n\mathbf{q}}|J|\Phi_{m\mathbf{q}}\rangle =0,
\pm 1$ ($+1$ for $n=m$ and positive energy, $-1$ for $n=m$ and negative 
energy, $0$ for $n\neq m$ or $\Delta n_{\mathbf{k}\mathbf{q}}^{ss'}=0$) 
and get the detail expression of the inverse transformation 
$\bar{U}^{-1}$. (For more details, please see the supplementary 
materials.).

\emph{Emitter SPP coupling}--Once getting $\bar{U}^{-1}$, we can 
calculate the interaction between quasi particles with other systems. 
As an example, suppose we have a emitter (a hydrogen like molecule) 
localized at $z=z_0,\ \mathbf{r}=0$ 
(Figure 1a). The plane wave functions $e^{i\mathbf{k}\cdot \mathbf{r}}$ 
and the discrete localized states $\phi_i(\mathbf{x})$ of emitter form 
a complete set, so we can expand
$\hat{\Psi}(\mathbf{x})=\hat{\Psi}_{2D}(\mathbf{x})+\hat{\Psi}_{et}
(\mathbf{x})
=1/\sqrt{S}\sum_{\mathbf{k}}e^{i\mathbf{k}\cdot 
\mathbf{r}}\psi(z)\Psi_{\mathbf{k}}^{\dag}+\sum_i
\phi_i(\mathbf{x})\hat{\phi}^{\dag}_i$,
here $\hat{\phi}_i^{\dag}$ is the $i^{th}$ creation operator 
of emitter. Note that the emitter's wave function is well localized, 
one reasonable assumption is its total electron number is conserved when 
interacting with SPPs. So, we only keep the emitter electron conserved 
terms in the total Hamiltonian which has three parts $\hat{H}=\hat{H}_
{et}+\hat{H}_{ee}+\hat{H}_{2D}$, where the last term $\hat{H}_{2D}$ is 
the Hamiltonian of 2DEG interacting with light defined 
previously, $\hat{H}_{et}$ is the emitter Hamiltonian under vector field 
$\hat{\mathbf{A}}(\mathbf{x})$ and emitter potential $V_{et}(\mathbf{x})$. 
The last term $\hat{H}_{ee}$ is electron electron interaction Hamiltonian, 
which can be expressed as the emitter electron interacting with fields 
generated by the 2DEG. Now 
\begin{align}
\hat{H}_{et}=&\int \hat{\Psi}_{et}^{\dag}
(\mathbf{x})[\frac{1}{2m^*}(\hat{p}+\frac{e}{c}\hat{A}
(\mathbf{x}))^2+V_{et}(\mathbf{x})]\hat{\Psi}(\mathbf{x})d\mathbf{x}\nonumber\\
\hat{H}_{ee}=&\int\int\hat{\Psi}_{et}^{\dag}(\mathbf{x})\hat{V}_{eff}
(\mathbf{x},\mathbf{x}')\hat{\Psi}_{et}(\mathbf{x}')d\mathbf{x}
d\mathbf{x}'.
\end{align}
The effective potential $\hat{V}_{eff}(\mathbf{x},\mathbf{x}')$ contains 
two parts, one originates from the direct Coulomb interaction $\delta
(\mathbf{x}-\mathbf{x}')\hat{V}_{ec}(\mathbf{x})$, and the other comes 
from electron exchanging effect $\hat{V}_{ex}(\mathbf{x},\mathbf{x}')$, 
where
\begin{align}
&\hat{V}_{ec}(\mathbf{x})=\int V(|\mathbf{x}-\mathbf{x}''|)
\hat{\Psi}_{2D}^{\dag}(\mathbf{x}'')
\hat{\Psi}_{2D}(\mathbf{x}'')d\mathbf{x}''\nonumber\\
&\hat{V}_{ex}(\mathbf{x},\mathbf{x}')=-V(|\mathbf{x}-\mathbf{x}'|)
\hat{\Psi}_{2D}^{\dag}(\mathbf{x}')\hat{\Psi}_{2D}(\mathbf{x}).
\end{align}

It should be emphasized here that, we want to derive the interaction 
between emitter and SPPs, but not the interaction between emitter with 
the total 2DEG. To do so, we can utilize the inverse 
transformation and only keep the contribution from SPPs in 
$\hat{V}_{eff}(\mathbf{x},\mathbf{x}')$. For the direct Coulomb potential, 
we have $\hat{V}^{spp}_{ec}(\mathbf{x})=\sum e^{-i\mathbf{q}\cdot
\mathbf{r}-q|z|}L_{\mathbf{q}}\hat{Q}_{\mathbf{q}}^{\dag}+h.c.$, 
where $L_{\mathbf{q}}$ is the normalization factor of wave function 
$\Phi_{\mathbf{q}}$, the superscript $^{spp}$ means only considering 
the contribution of SPPs. 

It is convenient to study the interaction through electric field in 
dipole approximation, and the contribution from $\hat{V}^{spp}_{ec}$ 
reads 
\begin{align}
&\hat{\mathbf{E}}^s(\mathbf{x})=\frac{1}{e}\nabla \hat{V}^{spp}_{ec}
(\mathbf{x})\nonumber\\
=&\frac{1}{e}\sum_{\mathbf{q}}(-i\mathbf{q},-q)e^{-i\mathbf{q}\cdot
\mathbf{r}-qz}L^{*}_{\mathbf{q}}\hat{Q}_{\mathbf{q}}^{\dag}+h.c. 
\end{align}
Similarly, we can calculate the vector field $\hat{\mathbf{A}}(\mathbf{x
})$, and its contribution to electric field reads
\begin{align}
\hat{E}_{\bot}^r(\mathbf{x})=&\frac{1}{e}
\sum_{\mathbf{q}}e^{-i\mathbf{q}\cdot\mathbf{r}}q(e^{-qz}-e^{-q'z})L^*_
{\mathbf{q}}\hat{Q}^{\dag}_{\mathbf{q}}+h.c.\nonumber\\
\hat{\mathbf{E}}_{\parallel}^r(\mathbf{x})=&\frac{1}{e}
\sum_{\mathbf{q}}e^{-i\mathbf{q}\cdot\mathbf{r}}
[i(\mathbf{q}e^{-qz}-\mathbf{q}'e^{-q'z}
)L^*_{\mathbf{q}}\nonumber\\
&+e^{-q'z}\Omega_{\mathbf{q}}\frac{\tilde{\mathbf{q}}}{q}
T^*_{\mathbf{q}}]\hat{Q}^{\dag}_{\mathbf{q}}+h.c.,
\end{align}
where $\tilde{\mathbf{q}}$ is an in-plane vector perpendicular to 
$\mathbf{q}$ ($\theta_{\tilde{\mathbf{q}}}=\theta_{\mathbf{q}}+\pi/2$, 
illustrated in Figure 1d), and $\mathbf{q}'$ means the magnitude of 
$\mathbf{q}$ reduced from $q$ to $q'=\sqrt{q^2-\Omega^2_{\mathbf{q}}/c^2}$, 
and $T_{\mathbf{q}}$ is another wave function normalization factor 
existing in transverse or longitudinal transverse hybrid modes. Finally, 
the total electric field reads
\begin{align}
&\hat{E}(\mathbf{x})=\hat{E}^s(\mathbf{x})+\hat{\mathbf{E}}^r(\mathbf{x})
=-\frac{1}{e}\sum_{\mathbf{q}}
e^{-i\mathbf{q}\cdot\mathbf{r}-q'z}\nonumber\\
&\times[(i\mathbf{q}',q)L^*_{\mathbf{q}}
+\Omega_{\mathbf{q}}\frac{\tilde{\mathbf{q}}}{q}
T^*_{\mathbf{q}}]\hat{Q}_{\mathbf{q}}^{\dag}+h.c..
\end{align}
\begin{figure}[t]
\begin{center}
\includegraphics[width=1\linewidth]{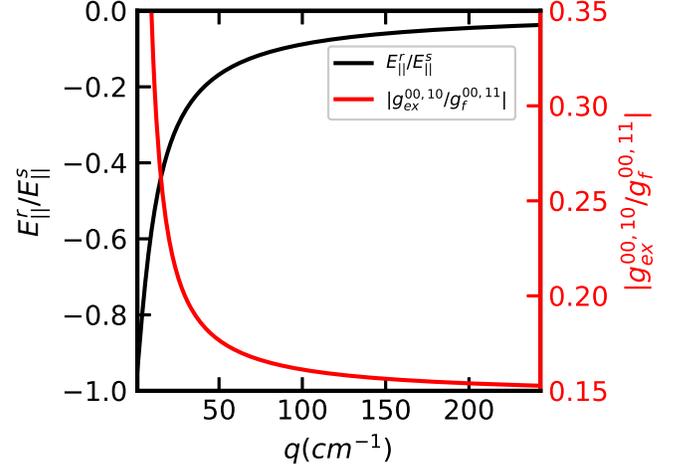} 
\end{center}
\caption{(Color online) The relative ratios $E^{r}_{||}/E^{s}_{||}$ at 
$z=0$ (black line) and $g^{ex}_{00,11}/g^f_{00,11}$ (red line), here 
the superscripts $^{r,s}$ mean radiation and static 
respectively, the subscript $_{||}$ means the in-plane component.}
\end{figure}
In Fig. 3, we've plotted the relative ratio of radiation to static 
electric field (the parallel constituent) as a function of $q$ at $z=0$. 
It is clearly that when $q$ is less that $100cm^{-1}$, the radiation 
component is big enough and can't be ignored. Because their vertical 
parts have the same magnitude but opposite sign at $z=0$, so we don't put 
it in the figure.

Now, the interaction Hamiltonian through electric field $\hat{H}_{int}^f
=-\int\hat{\mu}(\mathbf{x})\cdot\hat{\mathbf{E}}(\mathbf{x})d\mathbf{x}$  
reads 
\begin{align}
\hat{H}_{int}^f=&e\int\hat{\Psi}_{et}^{\dag}(\mathbf{x})
[\mathbf{x}\cdot\hat{\mathbf{E}}(\mathbf{x})]\hat{\Psi}_{et}(\mathbf{x})
d\mathbf{x}\nonumber\\
\approx&\sum_{ij\mathbf{q}}g_f^{ij}
(\mathbf{q})\hat{\phi}^{\dag}_i\hat{\phi}_j\hat{Q}^{\dag}_{\mathbf{q}}
+h.c.,
\end{align}
with the interaction strength 
\begin{align}
g_f^{ij}(\mathbf{q})=-i(\mathbf{q}'\cdot\mathbf{d}^{||}_{ij}
+qd^{\bot}_{ij})L^*_{\mathbf{q}}-\frac{\Omega_{\mathbf{q}}}{q}
\tilde{\mathbf{q}}\cdot\mathbf{d}^{||}_{ij}T^*_{\mathbf{q}},
\end{align}
where $\mathbf{d}_{ij}=\int\phi_i^*(\mathbf{x})\mathbf{x}
\phi_j(\mathbf{x}) d\mathbf{x}$ is the transition dipole 
moment, the superscript $^{||},^{\bot}$ means the parallel and 
vertical parts.

When the emitter and 2DEG get close to each other and their 
wave functions overlap in z direction, $\hat{V}_{ex}(\mathbf{x},
\mathbf{x}')$ arising from the electron exchanging becomes important, 
after some similar calculations, we 
find the exchange Hamiltonian $\hat{H}_{int}^{ex}$ has the same form of 
Eq. (10) with the interaction strength
\begin{align}
&g^{ij}_{ex}(\mathbf{q})=\frac{q\lambda_fL^*_{\mathbf{q}}}{8\pi^3}
\eta^{ij}_{\mathbf{q}},
\end{align}
where $\lambda_f$ is the Fermi wavelength of 2DEG, and 
$\eta^{ij}_{\mathbf{q}}$ is defined in the supplementary material. 
The exchange strength or $\eta^{ij}_{\mathbf{q}}$ is quite complicate 
and should be numerically calculated, but for a simple situation 
that a longitudinal SPP interacts with a 2D hydrogen like molecule, it 
has analytical results (Even in this simplified case, the 
expression is quite complicate. For the detail calculation, please see 
the supplementary materials). 

Note that the field interaction strength $g^{f}_{ij}(\mathbf{q})$ now 
equals $-i\mathbf{q}'\cdot\mathbf{d}_{ij}L^*_{\mathbf{q}}$, and we can 
compare it with the exchange strength by calculate the ratio 
$|g_{ex}^{ij}(\mathbf{q})/g^{f}_{ij}(\mathbf{q})|=\eta_{ij}(\lambda_f/a
_0)/(8\pi^3)q\lambda_f/(\mathbf{q}'\cdot\mathbf{d}_{ij})$. When $q$ isn't 
very small, $q'$ approx $q$ and these two interaction strengths are 
comparable. When $q\rightarrow 0$, because $q/q'\rightarrow \infty$, we 
can conclude that the exchange effect plays the most important role at 
small $q$. To provide a perspicuous and direct impression, we've plotted 
$|g_{ex}^{00,10}(\mathbf{q})/g_{f}^{00,11}(\mathbf{q})|$ for the 
transitions from ground state $(00)$ to the first $(10)$ and second 
$(11)$ excited states in Figure 3 (red line). Because the two strengths 
have different selection rules (For example, $g_{ex}^{00,10}(\mathbf{q})$ 
is none zero, while the transition dipole moment $\mathbf{d}_{00,10}=0$), 
and $g_{ex}^{00,11}(\mathbf{q})$ is quite complicate, we've calculated 
$|g_{ex}^{00,10}(\mathbf{q})/g_{f}^{00,11}(\mathbf{q})|$ instead of 
$|g_{ex}^{00,11}(\mathbf{q})/g_{f}^{00,11}(\mathbf{q})|$. It is clearly 
that it diverges at $q=0$, and approx $15\%$ at large $q$.

\emph{Conclusion.}--We have developed a quantization scheme for 
the collective excitations of 2DEGs when consider photon fields. The 
quasi particle energy and wave function have been derived, 
and we've calculated the inverse operator transformation which interprets 
photons and electron density oscillations with quasi particles. With the 
help of such transformation, the interaction between SPPs and molecule 
emitters are studied. We find when they are closely placed, a new type 
of interaction originating from exchanging electron emerges, which has 
different selection rules compared to the traditional interaction and has 
comparable strength at large $\mathbf{q}$, and plays the most important 
role at small momentum ($\mathbf{q}\lesssim 100cm^{-1}$). Finally, we 
want to emphasize here that this method can also be applied to other type 
emitters such as excitons and quantum dots, and other 2DEGs beyond TI 
surface states, and one can utilize it to quantize other excitations 
under RPA. 

We thank Prof. Shunping Zhang and Prof. Peter Nordlander for the helpful 
discussions. This work is supported by the Ministry of Science and Technology 
(Grant 2015CB932401), the National Key R\&D Program of China (Grant
2017YFA0205800, 2017YFA0303402), the National Natural Science Foundation of 
China (Grant 11344009).
 \bibliography{tpp1} 

\end{document}